\documentclass[epj]{svjourmod}
\usepackage{graphics,graphicx}

\newcommand{\rEs}{\langle r_{E,s}^2 \rangle}
\newcommand{\rMs}{\langle r_{M,s}^2 \rangle}
\newcommand{\Order}[1]{\mathcal{O}#1}
\newcommand{\Lagr}{\mathcal{L}}
\begin{document}
\title{Strange form factors and Chiral Perturbation Theory}
\author{Bastian Kubis
\thanks{\emph{Present address:} 
  HISKP (Th), Universit\"at Bonn, Nussallee 14-16, D-53115 Bonn, Germany
}
}    
\institute{Institut f\"ur theoretische Physik, Universit\"at Bern,
Sidlerstrasse 5, CH-3012 Bern, Switzerland
}
\date{
}
%
\abstract{
We review the contributions of Chiral Perturbation Theory to the
theoretical understanding or not-quite-yet-understanding of the nucleon
matrix elements of the strange vector current.
} 
\maketitle
%


\section{Introduction}
\label{intro}

Chiral Perturbation Theory (ChPT), as the low-energy effective field
theory of the Standard Model, ought to be predestined to describe the 
response of the nucleon to the strange vector current: 
it incorporates all symmetry constraints of the fundamental theory and
contains the degrees of freedom relevant at low energies, Goldstone
bosons (pions and kaons) as well as matter fields (nucleons).
In Sect.~\ref{sec:chpt}, we shall point out
some specific aspects of ChPT that are essential for
the following discussion.
We shall reiterate in Sect.~\ref{sec:ms+res} why symmetry
considerations alone are not sufficient for ChPT to be predictive for 
the leading moments of the strange form factors, the strange magnetic
moment $\mu_s$ and the strange electric radius $\rEs$.
A low-energy theorem for the strange \emph{magnetic} radius $\rMs$
is presented in Sect.~\ref{sec:rms}, its usefulness and limitations
are discussed. 
Some alternative regularization schemes for loop diagrams have been
suggested in order to achieve improved convergence behavior of the
chiral series, we shall say a few words on these in Sect.~\ref{sec:reg}.
Finally, a conclusion is given in Sect.~\ref{sec:concl}.


\section{Some aspects of Chiral Perturbation Theory}
\label{sec:chpt}


\subsection{Goldstone bosons and counterterms}
\label{sec:counter}

ChPT~\cite{gl} systematically exploits the far-reaching consequences of chiral
symmetry:  it dictates the appearance of (pseudo) Goldstone bosons
($\pi$, $K$, $\eta$) and tightly constrains their interaction with
each other as well as with external currents and matter fields.  
As a consequence, the Goldstone boson dynamics is completely
calculable, while the influence of heavier states can be parameterized,
at low energies, by polynomials.  
The polynomial coefficients (called low-energy constants) are not
numerically known \textit{a priori}, but are still far from arbitrary, as
they potentially interrelate many different observables.  A particularly famous
example is the quark mass expansion of the pion mass:
\begin{equation}\label{GMOR}
M_\pi^2 = 2B\hat{m}
- \frac{\bar{\ell}_3}{32\pi^2F_\pi^2} \bigl[2B\hat{m}\bigr]^2 
+ \Order\left(\hat{m}^3\right)~~,
\end{equation}
where $\hat{m}=(m_u+m_d)/2$.  
The first term on the right constitutes the well-known
Gell-Mann--Oakes--Renner relation.  The correction of order
$\hat{m}^2$ is proportional to the low-energy constant
$\bar{\ell}_3$ that can be determined from
$\pi\pi$ scattering.  Given such independent experimental
information, the relative size of the contributions to $M_\pi^2$
linear and quadratic in the quark masses is known~\cite{condensate}.


\subsection{Power counting}
\label{sec:power}
In the Goldstone boson sector of ChPT, Lorentz invariance
dictates that only even powers of momenta appear in the effective
Lagrangian.  The typical low-energy expansion parameters therefore are 
$M_\pi^2/\Lambda_\chi^2 \approx 0.02$ (for chiral SU(2)) or 
$M_K^2/\Lambda_\chi^2 \approx 0.2$ (for chiral SU(3)), where the
chiral symmetry breaking scale is $\Lambda_\chi=4\pi F_\pi\approx 1160$~MeV.
In contrast, due to the appearance of spin, there are also odd powers
of momenta in the effective pion-nucleon \linebreak  Lagrangian,
such that the convergence order-by-order is \linebreak  markedly slower,
typical expansion parameters are \linebreak $M_\pi/m_N \approx 0.15$ in SU(2) or 
$M_K/m_N\approx 0.5$ for SU(3). 
Obviously, there is no way we can expect ChPT to work as well for the
baryon sector with strangeness as it does for, say, $\pi\pi$ scattering.

It is important to understand the ``generic'' chiral orders of the
lowest moments in the  nucleon vector form factors, i.e.\ the orders
at which polynomial contributions occur.
The low-energy expansion of the strange Sachs form factors is given by 
\begin{eqnarray*}
G_{E,s}\left(Q^2\right) &=& 
\underbrace{\phantom{\frac{1}{6}}\hskip -0.3cm Q_s}_{\Order(p)}
  -\;\underbrace{\frac{1}{6} \rEs \,Q^2}_{\Order(p^3)} \,\,+ \,\ldots ~~,\\
G_{M,s}\left(Q^2\right) &=&
\underbrace{\phantom{\frac{1}{6}}\hskip -0.3cm \mu_s}_{\Order(p^2)}
  \!-\;\underbrace{\frac{1}{6} \rMs \,Q^2}_{\Order(p^4)} \,+ \,\ldots ~~,
\end{eqnarray*}
where one has $Q_s=0$ due to gauge invariance for the strange charge. 
We note in particular that polynomial contributions to the radius
terms $\rEs$, $\rMs$ appear at leading and subleading one-loop order,
respectively. 


\section{Why ChPT cannot predict \boldmath{$\mu_s$} and \boldmath{$\rEs$}}
\label{sec:ms+res}
The reason why it is impossible to predict $\mu_s$ and $\rEs$ from
first principles in ChPT
was identified several years ago~\cite{ito}.  There are three
independent diagonal vector currents in SU(3),
\begin{equation}
J_\mu^{(i)} = \bar{q} \frac{\lambda^i}{2}\gamma_\mu q ~~, \qquad
i=3,\,8,\,0,
\end{equation}
where $\lambda^{3/8}$ are the usual Gell-Mann matrices and
$\lambda^0=\sqrt{2/3}\;{\rm diag}(1,1,1)$.  
The three are proportional to the iso\-vector and isoscalar
electromagnetic currents and the \linebreak baryon number current, respectively.
The electromagnetic and the strangeness currents are linear
combinations of these, 
\begin{equation}
J_\mu^{\rm EM} = 
  J_\mu^{(3)} + \frac{1}{\sqrt{3}}J_\mu^{(8)} ~~,\quad
J_\mu^s = 
  \sqrt{\frac{2}{3}} J_\mu^{(0)} - \frac{2}{\sqrt{3}} J_\mu^{(8)}~~, 
\end{equation}
i.e.\ the response to one component of the strangeness current, the
baryon number, is going to be completely independent of what we know
from electromagnetic probes.  In the effective Lagrangian language,
this means that wherever matrix elements of the electromagnetic
current depend on low-energy constants, there will be a \emph{new,
  independent} constant for the strangeness current.  As an example,
consider the leading terms contributing to the magnetic moments,
\begin{equation}
\Lagr^{(2)} = \frac{b_6^{D/F}}{8m_N}
  \langle \bar{B} \sigma^{\mu\nu} \left[ F_{\mu\nu}^+, B \right]_\pm \rangle
+ \frac{b_6^0}{8m_N} 
  \langle \bar{B} \sigma^{\mu\nu} B \rangle \langle F_{\mu\nu}^+ \rangle ~.
\end{equation}
The constants $b_6^{D/F}$ can be fitted alternatively to the magnetic
moments of proton and neutron or to all octet moments, but $b_6^0$
only appears in the strange magnetic moment.
The same pattern emerges for all low-energy constants, therefore
instead of predicting the strange vector form factors, ChPT can only
adjust its constants in order to reproduce experimental findings.


\section{The strange magnetic radius}
\label{sec:rms}
Besides being a quantity of interest in its own right, the strange
magnetic radius $\rMs$ is also of high importance for the experimental
determination of the strange magnetic \emph{moment} $\mu_s$ for the
following reason:  Experimental measurements of $G_{M,s}(Q^2)$ always
have to be performed at finite, non-vanishing $Q^2$, therefore one
needs to extrapolate to $G_{M,s}(0)=\mu_s$ \cite{HMS}. 

\subsection{A low-energy theorem for \boldmath{$\rMs$}}
\label{sec:rmslet}

\begin{figure} 
\begin{center}
\resizebox{0.25\textwidth}{!}{
  \includegraphics{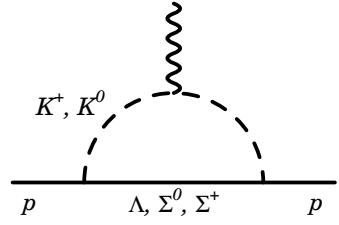}
}
\end{center}
\caption{Diagram generating the leading contribution to $\rMs$}
\label{fig:loop}
\end{figure}
\begin{figure} 
\begin{center}
\resizebox{0.3\textwidth}{!}{
  \includegraphics{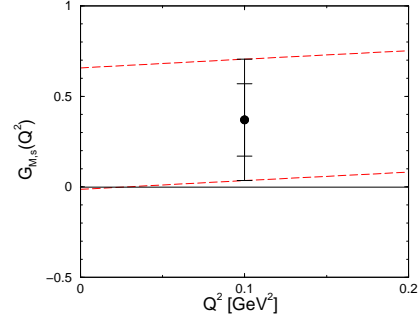}
}
\end{center}
\caption{Extrapolation of the SAMPLE value $G_{M,s}(Q^2)$ to $\mu_s$}
\label{fig:extp3}
\end{figure}

After the pessimistic finding of the last section, how can there
possibly be any low-energy theorem for any strange vector form factor?  
The answer is, through leading non-analytic loop effects.  
The diagram of order $\Order(p^3)$ displayed in Fig.~\ref{fig:loop}
generates a contribution to the strange magnetic radius according
to~\cite{HMS} 
\begin{equation}\label{rmsp3}
\rMs^{(3)} = - \frac{5D^2-6DF+9F^2}{48\pi F_K^2}\frac{m_N}{M_K} 
+ \Order(M_K^0) ~~. 
\end{equation}
As a low-energy constant can only contribute to $\rMs$ at
the next order, generating a term of $\Order(M_K^0)$, the term in
(\ref{rmsp3}) is (at least formally) dominant.  
All the masses and coupling constants in (\ref{rmsp3}) are
known, hence we have a parameter-free prediction for $\rMs$, 
\begin{equation}\label{rmsp3num}
\rMs = -0.115~{\rm fm}^2 ~~.
\end{equation}
We can use the leading chiral prediction for $\rMs$ 
to extrapolate from the SAMPLE result \cite{sample} 
$G_{M,s}(0.1\,{\rm GeV}^2) = 0.37\pm0.20\pm0.26\pm0.07$
to the strange magnetic moment,
\begin{equation}
\mu_s = 0.32\pm0.20\pm0.26\pm0.07 ~~.
\end{equation}
This extrapolation is visualized in Fig.~\ref{fig:extp3}.
\begin{figure}[htb]
\begin{center}
\resizebox{0.3\textwidth}{!}{
  \includegraphics{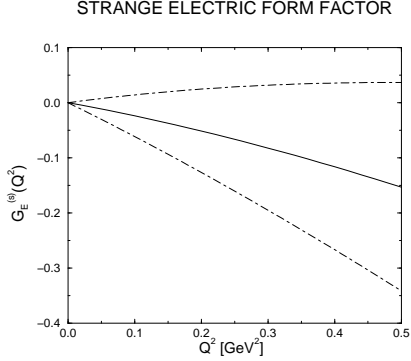}
}
\end{center}
\caption{$Q^2$--dependence of $G_{E,s}$ at $\Order(p^3)$.  The figure
  is taken from~\cite{HKM}}
\label{fig:GEs}
\end{figure}
Furthermore, in~\cite{HKM} this was combined with a second
experimental result, the HAPPEX measurement~\cite{happex} for
$G_{E,s}+0.39G_{M,s}$, in order to fix both unknowns in the
$\Order(p^3)$ representation of the strange form factors (for 
$\mu_s$ and $\rEs$) and predict also the $Q^2$--dependence of the
strange electric form factor, see Fig.~\ref{fig:GEs}.
We note from that figure that even over a large $Q^2$ range, the form
factor looks essentially linear and displays only very little
curvature.  The reason for this is that the closest infrared singularity
in this representation is the $K\bar{K}$ cut, which has a threshold
far away from the physical region, and that counterterms providing
curvature in the polynomial part are suppressed to two-loop order.

\subsection{Stability of the low-energy theorem}
\label{sec:rMsp4}
\begin{figure}
\begin{center}
\includegraphics[height=1.65in,angle=0,clip=true]{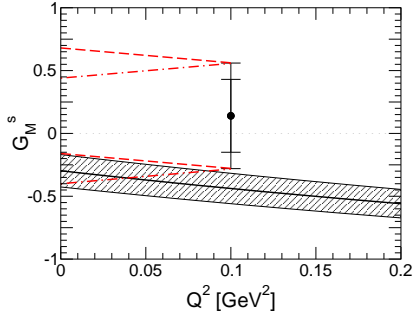}
\end{center}
\caption{Extrapolation of the SAMPLE value $G_{M,s}(Q^2)$ to $\mu_s$,
  using the $\Order(p^4)$ value for $\rMs$. 
  The grey band indicates a dispersion-theoretical prediction.  The
  figure is taken from~\cite{hammer}} 
\label{fig:extp4}
\end{figure}

Even though the low-energy theorem (\ref{rmsp3}) is strictly a
\emph{theorem}, it is an important problem to study its stability when
subject to higher-order corrections.  
A calculation to $\Order(p^4)$ has been performed~\cite{hammer} (see
also~\cite{diss}), from which we only quote the numerical result,
\begin{equation}\label{rmsp4}
\rMs^{(4)} = - \bigl(0.04 + 0.3\,b_s^r\bigr)~{\rm fm}^2 ~~,
\end{equation}
where the low-energy constant $b_s^r$ is expected to be of order 1.  
We note that the corrections to the central value as compared to
(\ref{rmsp3num}) are sizeable, and that the unknown constant
induces a theoretical uncertainty in the extraction 
of $\mu_s$ from experiment.  The corresponding plot is shown in
Fig.~\ref{fig:extp4} (taken from~\cite{hammer}). 


\section{Variants of loop regularization}
\label{sec:reg}

We have seen that the low-energy theorem for $\rMs$ is given by a
leading-order kaon-loop effect, but that the next-to-leading order
corrections are sizeable.  We therefore want to discuss here two
alternative schemes to evaluate loops that may resum these corrections
more efficiently, or may give a fairer estimate of these contributions
altogether.


\subsection{Infrared regularization}
\label{sec:ir}

All loop results discussed so far have been calculated in what is
called Heavy Baryon Chiral Perturbation Theory.  In this formalism,
the nucleons are treated as heavy, non-relativistic fields, for which
the relativistic corrections suppressed by powers of $1/m_N$ can be
calculated systematically.  This method has the advantage that loop
diagrams always follow naive power counting rules and that leading
loop calculations are technically simple, but the downside that the
analytic structure is sometimes distorted even in the low-energy
region, and that in practice $1/m_N$ corrections are awkward to calculate.

\begin{figure} 
\begin{center}
\resizebox{0.485\textwidth}{!}{
  \includegraphics{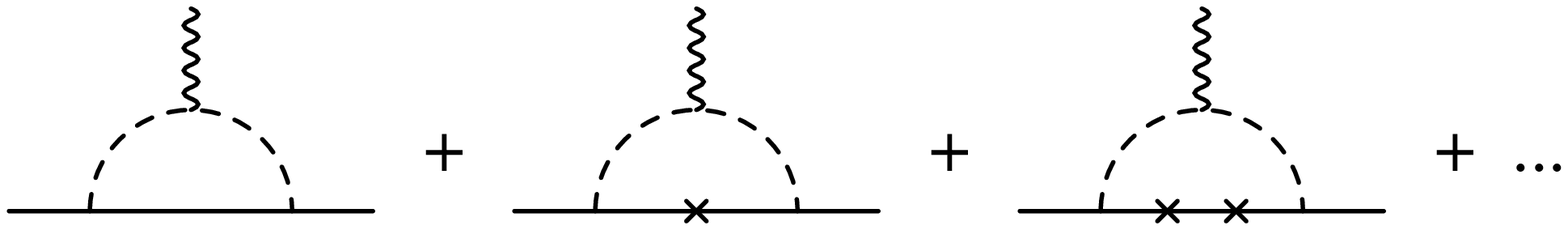}
}
\end{center}
\caption{Resummation of relativistic corrections in the infrared
  regularization scheme.  The crosses denote  $1/m_N$
  insertions in the nucleon propagator}
\label{fig:ir}
\end{figure}

One alternative scheme that has been suggested to overcome these
shortcomings is ``infrared regularization'' \cite{becher}, which is a
variant of dimensional regularization that preserves power counting
also in relativistic baryon ChPT.  In this scheme, all the $1/m_N$
corrections are automatically resummed as depicted schematically in
Fig.~\ref{fig:ir}.

It has been shown in~\cite{irff} that in some cases particularly
sensitive to relativistic ``recoil effects'' (in~\cite{irff}: the neutron
electric form factor), this relativistic resummation leads to a much
improved convergence behavior.
It is therefore interesting to see what happens to the
strange magnetic radius in this scheme, even beyond $\Order(p^4)$. 
We find the following \emph{partial} corrections of $\Order(p^5)$ \cite{diss}:
\begin{eqnarray}
\Delta \rMs^{(5)} &=& - \frac{35(5D^2\!-\!6DF\!+\!9F^2)}{384\pi F_K^2}
\frac{M_K}{m_N}
+\Order\biggl(\!\frac{\Delta m_{\Lambda/\Sigma}}{M_K}\!\biggr) \nonumber\\
&=& + 0.053~{\rm fm}^2 ~~,
\end{eqnarray}
where the additional terms proportional to $\Delta m_{\Lambda/\Sigma}
= m_{\Lambda/\Sigma} -m_N$ have been included in the numerical value.
Although these expressions are not complete (there will also be two-loop
contributions), they are non-analytic in
the quark masses and therefore not modified by counterterms.
Numerically, they turn out to be large, about half the size of the
leading $\Order(p^3)$ value.  
This supports once more the conclusion that, unfortunately, the
low-energy theorem for $\rMs$ at $\Order(p^3)$ is numerically not very
reliable. 


\subsection{Cutoff regularization}
\label{sec:cutoff}

Various attempts at cutoff regularization have been undertaken in the
context of ChPT.  We do not try to give an overview of these, but instead
concentrate on one specific approach~\cite{cutoff} in which $\rMs$ has
explicitly been considered~\cite{umass}.  The idea is that the nucleon
has an intrinsic size, and that ChPT can only reliably predict the
$\pi$ or $K$ fluctuations that are long-ranged on this scale.  Dimensional
regularization does not separate these different ranges of momenta in
the loop integration, such that it might be more efficient to employ a
cutoff in order to reduce unrealistically strong short-distance
effects.

\begin{figure} 
\begin{center}
\resizebox{0.3\textwidth}{!}{
  \includegraphics{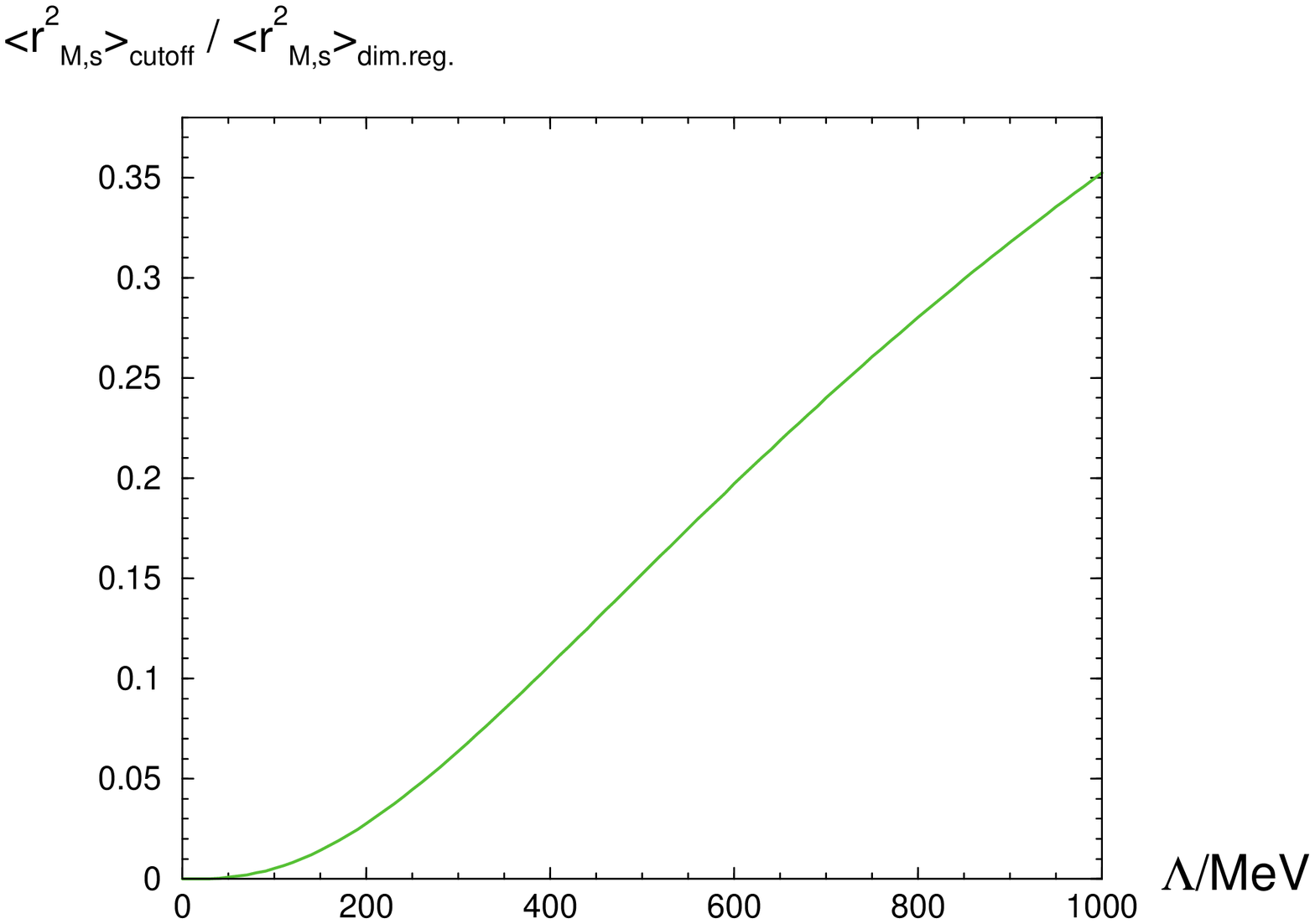}
}
\end{center}
\caption{Cutoff dependence of $\rMs$}
\label{fig:tobias}
\end{figure}

The method in~\cite{cutoff} amounts to the use of a $KNN$ form
factor instead of a constant vertex.
In~\cite{umass}, a dipole form  has been chosen, which leads to
the analytic result
\begin{eqnarray}
\rMs_{\rm cutoff}&=&\rMs_{\rm dim.reg.}
  \times X\bigl({\textstyle \frac{\Lambda}{M_K}}\bigr) ~~, \\
X(x) &=& \frac{x^3(x^2+\frac{14}{5}x+1)}{(x+1)^{5}} ~~.\nonumber
\end{eqnarray}
This cutoff dependence is displayed in Fig.~\ref{fig:tobias}.  
It is obvious that for cutoffs of the order of
$\Lambda=400\ldots600$~MeV,  $\rMs_{\rm cutoff}$ is sizeably reduced
compared to the dimensional regularization result.

Despite the host of plausible arguments in favor of cutoff schemes,
there are also some drawbacks:
\begin{enumerate}
\item The cutoff is an additional parameter that can only, at best, be
  estimated.
\item It presents a deviation from the strict effective Lagrangian approach in
  ChPT, as a consequence of which gauge invariance usually has to be
  cured ``by hand'', and chiral symmetry is by no means
  guaranteed.\footnote{This problem has recently been addressed
  in~\cite{scherercutoff}.} 
\item The cutoff upsets the analytic structure of the form factors: 
  it produces additional unphysical poles and/or cuts.  This means in
  particular that is is unclear how a marriage with dispersion
  relations (see e.g.~\cite{hammer} and references therein) might be achieved.
\item Finally, the reduction of the numerical value of $\rMs$ must not
  be confused with the inclusion of higher order corrections as done in the
  infrared regularization scheme.  The problem of large 
  $1/m_N$ corrections in chiral SU(3) is by no means solved here.
\end{enumerate}


\section{Summary and conclusions}
\label{sec:concl}
The matrix elements of the strange vector current 
seem to remain elusive quantities for a description in Chiral
Perturbation Theory.
They are problematic as  Goldstone boson
dynamics is not overly dominant here, and in most cases, unknown
low-energy constants appear at leading order.  
We have pointed out that there is an exception, the strange
magnetic moment for which a parameter-free low-energy theorem exists.
However, higher order corrections are found to
be  sizeable, such that the low-energy theorem is very unstable once
these are taken into account.  Also cutoff calculations indicate 
a smaller absolute size of $\rMs$.  
The role of ChPT remains however to aid to interrelate more data, as
they become available~\cite{a4}.


\bigskip \noindent
\textit{Acknowledgements.}
I would like to thank the organizers of PAVI 04 for the great
opportunity to participate in this workshop, and for the wonderful organization
of the whole event.  
I am grateful to T.R.~Hemmert and U.-G.~Mei{\ss}ner for the
fruitful collaboration that originally introduced me to this subject,
and to various colleagues for interesting discussions that deepened my
understanding of the field, in particular H.-W. Hammer,
M.J. Ramsey-Musolf, J.F. Donoghue, B.R. Holstein, T. Huber, and
A. Ro{\ss}.   
This work was supported in part by RTN, BBW-Contract 
No.~01.0357, and EC-Contract HPRN--CT2002--00311 \linebreak (EURIDICE).


\end{document}